%% file: LIGO-P1500217_GW150914_Rates_Supplement.tex
\documentclass[twocolumn,twocolappendix]{aastex6}

\pdfoutput=1

\usepackage{acronym}
\usepackage{amsmath}
\usepackage{graphicx}
\usepackage{hyperref}
\usepackage{lineno}
\usepackage{multirow}

\newcommand{\svnid}[1]{ } 

\input{macros.tex}

\renewcommand{\macro}[1]{{#1}}

\newcommand{\emcee}{\texttt{emcee}}

\newcommand{\xmin}{\ensuremath{x_\mathrm{min}}}

\newcommand{\firstevent}{GW150914}
\newcommand{\secondevent}{\SECONDMONDAY}

\newcommand{\avgVT}{\ensuremath{\left\langle VT \right\rangle}}

\newcommand{\dd}{\ensuremath{\mathrm{d}}}
\newcommand{\diff}[2]{\ensuremath{\dfrac{\dd {#1}}{\dd {#2}}}}

\newcommand{\gstlal}{\texttt{gstlal}}
\newcommand{\pycbc}{\texttt{pycbc}}

\usepackage{color}
\usepackage{graphicx}

\newcommand{\letseccounts}{2.1}
\newcommand{\letsecmassdistribution}{3}
\newcommand{\letsecrates}{2.2}

\newcommand{\leteqaveragespacetimevolume}{(15)}
\newcommand{\leteqlikelihood}{(10)}
\newcommand{\leteqpfore}{(7)}
\newcommand{\leteqtwopopfgmcrate}{(1)}
\newcommand{\leteqtwopoplikelihood}{(3)}
\newcommand{\leteqtwopopposterior}{(5)}

\begin{document}

\title{Supplement: The Rate of Binary Black Hole Mergers Inferred from Advanced
  LIGO Observations Surrounding \firstevent{}}

\AuthorCallLimit=1
\fullcollaborationName{The LIGO Scientific Collaboration and Virgo
  Collaboration}
\include{authors}

\begin{abstract}
  Supplemental information for a Letter reporting the rate of \ac{BBH}
  coalescences inferred from \OBSDAYS{} of coincident Advanced LIGO
  observations surrounding the transient \ac{GW} signal \firstevent{}.
  In that work we reported various rate estimates whose 90\% \acp{CI}
  fell in the range \oneraterangetorulethemall{}.  Here we give
  details of our method and computations, including information about
  our search pipelines, a derivation of our likelihood function for
  the analysis, a description of the astrophysical search trigger
  distribution expected from merging \acp{BBH}, details on our
  computational methods, a description of the effects and our model
  for calibration uncertainty, and an analytic method of estimating
  our detector sensitivity that is calibrated to our measurements.
\end{abstract}

\maketitle

\acrodef{BH}[BH]{black hole}
\acrodef{BBH}[BBH]{binary black hole}
\acrodef{CI}[CI]{credible interval}
\acrodef{CL}[CL]{credible level}
\acrodef{CBC}[CBC]{compact binary coalescence}
\acrodef{EOB}[EOB]{effective one body}
\acrodef{FAP}{false alarm probability}
\acrodef{FAR}{false alarm rate}
\acrodef{GRB}{gamma ray burst}
\acrodef{GW}[GW]{gravitational wave}
\acrodef{KS}{Kolmogorov-Smirnov}
\acrodef{PE}{parameter estimation}
\acrodef{SNR}{signal-to-noise ratio}
\acrodefplural{SNR}{signal-to-noise ratios}

\acresetall{} 
The first detection of a \ac{GW} signal from a merging \ac{BBH} system
is described in \citet{GW150914-DETECTION}.  \citet{RatesLetter}
reports on inference of the local \ac{BBH} merger rate from
surrounding Advanced LIGO observations.  This Supplement provides
supporting material and methodological details for
\citet{RatesLetter}, hereafter referred to as the Letter.

\section{Search Pipelines}
\label{suppsec:search-description}

Both the \pycbc{} and \gstlal{} pipelines are based on matched
filtering against a bank of template waveforms.  See
\citet{GW150914-CBC} for a detailed description of the pipelines in
operation around the time of \firstevent{}; here we provide an
abbreviated description.  

In the \pycbc{} pipeline, the single-detector \ac{SNR} is re-weighted
by a chi-squared factor \citep{Allen:2004gu} to account for
template-data mismatch \citep{babak:2012zx}; the re-weighted
single-detector \acp{SNR} are combined in quadrature to produce a
detection statistic for search triggers.  

The \gstlal{} pipeline's detection statistic, however, is based on a
likelihood ratio \citep{Cannon2013,Cannon2015} constructed from the
single-detector \acp{SNR} and a signal-consistency statistic.  An
analytic estimate of the distribution of astrophysical signals in
multiple-detector \ac{SNR} and signal consistency statistic space is
compared to a measured distribution of single-detector triggers
without a coincident counterpart in the other detector to form a
multiple-detector likelihood ratio.

Both pipelines rely on an empirical estimate of the search background,
making the assumption that triggers of terrestrial origin occur
independently in the two detectors.  The background estimate is built
from observations of single-detector triggers over a long time
(\gstlal{}) or through searching over a data stream with one
detector's output shifted in time relative to the other's by an
interval that is longer than the light travel time between detectors,
ensuring that no coincident astrophysical signals remain in the data
(\pycbc{}).  For both pipelines it is not possible to produce an
instantaneous background estimate at a particular time; this drives
our choice of likelihood function as described in Section
\ref{suppsec:likelihood}.

The \gstlal{} pipeline natively determines the functions $p_0(x)$ and
$p_1(x)$ for its detection statistic $x$.  For this analysis a
threshold of $\xmin = 5$ was applied, which is sufficiently low that
the trigger density is dominated by terrestrial triggers near
threshold.  There were $M=15\,848$ triggers observed above this
threshold in the 17 days of observation time analyzed by \gstlal{}.

\begin{figure}
  \plotone{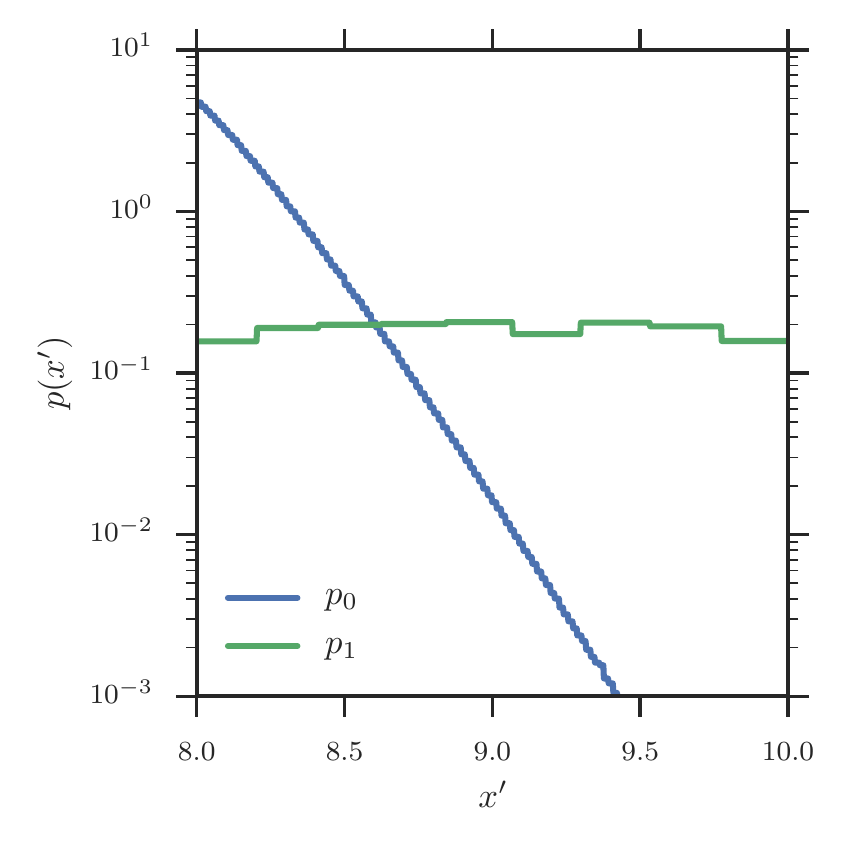}
  \caption{Inferred terrestrial ($p_0$; blue) and astrophysical
    ($p_1$; green) trigger densities for the \pycbc{} pipeline as
    described in Section \ref{suppsec:search-description}. }
  \label{fig:supp-p0p1}
\end{figure}

For \pycbc{}, the quantity $x'$ is the re-weighted \ac{SNR} detection
statistic.\footnote{When quoting pipeline-specific values we
  distinguish \pycbc{} quantities with a prime.}  We set a threshold
$\xmin' = \newsnrthresh{}$, above which $M' = 270$ triggers remain in
the search.  We use a histogram of triggers collected from
time-shifted data to estimate the terrestrial trigger density,
$p_0\left(x'\right)$, and a histogram of the recovered triggers from
the injection sets described in Section \letsecrates{} of the Letter
to estimate the astrophysical trigger density, $p_1\left(x'\right)$.
These estimates are shown in Figure \ref{fig:supp-p0p1}.  The
uncertainty in the distribution of triggers from this estimation
procedure is much smaller than the uncertainty in overall rate from
the finite number statistics (see, for example, Figure
\ref{fig:foreground-background}).  The empirical estimate is necessary
to properly account for the interaction of the various single- and
double-interferometer thresholds in the \pycbc{} search
\citep{GW150914-CBC}.  At high \ac{SNR}, where these thresholds are
irrelevant, the astrophysical triggers follow an approximately
flat-space volumetric density (see Section \ref{suppsec:universal}) of
\begin{equation}
  \label{eq:newsnr-foreground}
  p_{1}(x') \simeq \frac{3 \xmin'^3}{x'^4},
\end{equation}
but they deviate from this at smaller \ac{SNR} due to threshold
effects in the search.

For the \pycbc{} pipeline, a detection statistic
$x' \geq \loudnewsnr{}$ corresponds to an estimated search \ac{FAR} of
one per century.  

\section{Derivation of Poisson Mixture Model Likelihood}
\label{suppsec:likelihood}

In this section we derive the likelihood function in Eq.\
\leteqtwopoplikelihood{} of the Letter.  Consider first a search of the type
described in Section \ref{suppsec:search-description} over $N_T$
intervals of time of width $\delta_i$,
$\left\{i = 1, \ldots, N_T\right\}$.  Triggers above some fixed
threshold occur with an instantaneous rate in time and detection
statistic $x$ given by the sum of the terrestrial and astrophysical
rates:
\begin{equation}
  \label{eq:time-snr-rate-sum}
  \diff{N}{t \mathrm{d} x}(t,x) = R_0(t) p_0(x; t) + R_1(t) V(t)
  p_1(x; t),
\end{equation}
where $R_0(t)$ is the instantaneous rate (number per unit time) of
terrestrial triggers, $R_1(t)$ is the instantaneous rate density
(number per unit time per unit comoving volume) of astrophysical
triggers, $p_0$ is the instantaneous density in detection statistic of
terrestrial triggers, $p_1$ is the instantaneous density in detection
statistic of astrophysical triggers, and $V(t)$ is the instantaneous
sensitive comoving redshifted volume \citep[][see also Eq.\
\leteqaveragespacetimevolume{} of the Letter]{GW150914-ASTRO} of the
detectors to the assumed source population.  The astrophysical rate
$R_1$ is to any reasonable approximation constant over our
observations so we will drop the time dependence of this term from
here on.\footnote{The astrophysical rate can, in principle, also
  depend on redshift, but in this paper we assume that the \ac{BBH}
  coalescence rate is constant in the comoving frame.}  Note that
$R_0$ and $R_1$ have different units in this expression; the former is
a rate (per time), while the latter is a \emph{rate density} (per
time-volume).  The density $p_1$ is independent of source parameters
as described in Section \ref{suppsec:universal}.  Let
\begin{equation}
  \label{eq:dNdt}
  \diff{N}{t} \equiv \int \mathrm{d} x \, \diff{N}{t \mathrm{d} x} =
  R_0(t) + R_1 V(t).
\end{equation}

If the search intervals $\delta_i$ are sufficiently short, they will
contain at most one trigger and the time-dependent terms in Eq.\
\eqref{eq:time-snr-rate-sum} will be approximately constant.  Then the
likelihood for a set of times and detection statistics of triggers,
$\left\{ (t_j, x_j) | j = 1, \ldots, M \right\}$, is a product over
intervals containing a trigger (indexed by $j$) and intervals that do
not contain a trigger (indexed by $k$) of the corresponding Poisson
likelihoods
\begin{multline}
  \label{eq:segmented-likelihood}
  \mathcal{L} = \left\{ \prod_{j = 1}^M \diff{N}{t \mathrm{d} x}\left(
      t_j, x_j\right) \exp\left[ -\delta_j \diff{N}{t}\left( t_j
      \right) \right] \right\} \\ \times \left\{ \prod_{k = 1}^{N_T - M}
    \exp\left[ - \delta_k \diff{N}{t}\left( t_k \right)\right]\right\}
\end{multline}
(cf.\ \citet[][Eq.\ (21)]{Farr2015} or \citet[][Eq.\
(2.8)]{Loredo1995}).\footnote{There is a typo in Eq.\ (2.8) of
  \citet{Loredo1995}.  The second term in the final bracket is missing
  a factor of $\delta t$.}  Now let the width of the observation
intervals $\delta_i$ go to zero uniformly as the number of intervals
goes to infinity.  Then the products of exponentials in Eq.\
\eqref{eq:segmented-likelihood} become an exponential of an integral,
and we have
\begin{equation}
  \label{eq:continuous-likelihood}
  \mathcal{L} = \prod_{j=1}^M \left[ \diff{N}{t \mathrm{d} x}\left( t_j,
      x_j\right) \right] \exp\left[ -N \right],
\end{equation}
where 
\begin{equation}
  \label{eq:dNdt-integrated}
  N = \int \mathrm{d} t \, \diff{N}{t}
\end{equation}
is the expected number of triggers of both types in the total
observation time $T$.

As discussed in Section \ref{suppsec:search-description}, in our
search we observe that $R_0$ remains approximately constant and that
$p_0$ retains its shape over the observation time discussed here; this
assumption is used in our search background estimation procedure
\citep{GW150914-CBC}.  The astrophysical distribution of triggers is
universal (Section \ref{suppsec:universal}) and also time-independent.
Finally, the detector sensitivity is observed to be stable over our
\OBSDAYS{} of coincident observations, so $V(t) \simeq \mathrm{const}$
\citep{GW150914-CALIBRATION}.  We therefore choose to simply ignore
the time dimension in our trigger set.  This generates an estimate of
the rate that is sub-optimal (i.e.\ has larger uncertainty) but
consistent with using the full data set to the extent that the
detector sensitivity varies in time; since this variation is small,
the loss of information about the rate will be correspondingly small.
We \emph{do} capture any variation in the sensitivity with time in our
Monte-Carlo procedure for estimating $\avgVT{}$ that is described in
Section \letsecrates{} of the Letter.

If we ignore the trigger time, then the appropriate likelihood to use
is a marginalization of Eq.\ \eqref{eq:continuous-likelihood} over the
$t_j$.  Let 
\begin{multline}
  \label{eq:time-marg-like}
  \bar{\mathcal{L}} \equiv \int \left[\prod_j \dd t_j\right] \, \mathcal{L} \\ =
  \prod_j \left[ \Lambda_0 p_0\left( x_j \right) + \Lambda_1 p_1\left(
      x_j \right) \right] \exp\left[- \Lambda_0 - \Lambda_1 \right],
\end{multline}
where 
\begin{equation}
  \label{eq:L0-marg}
  \Lambda_0 p_0(x) = \int \dd t \, R_0(t) p_0\left( x; t \right),
\end{equation}
and
\begin{equation}
  \label{eq:L1-marg}
  \Lambda_1 p_1(x) = \int \dd t \, R_1 V(t) p_1\left( x; t \right),
\end{equation}
with 
\begin{equation}
  \label{eq:norm-L0L1}
  \int \dd x \, p_0(x) = \int \dd x \, p_1(x) = 1.
\end{equation}
If we assume that $R_1$ is constant in (comoving) time, and measure
$p_1(x)$ by accumulating recovered injections throughout the run as we
have done, then this expression reduces to the likelihood in Eq.\
\leteqtwopoplikelihood{} of the Letter.  A similar argument with an additional
population of triggers produces Eq.\ \leteqlikelihood{} of the Letter.

\subsection{The Expected Number of Background Triggers}
The procedure for estimating $p_0(x)$ in the \pycbc{} pipeline also
provides an estimate of the mean number of background events per
experiment $\Lambda_0$ \citep{GW150914-CBC}.  The procedure for
estimating $p_0$ used in the \gstlal{} pipeline, however, does not
naturally provide an estimate of $\Lambda_0$; instead \gstlal{}
estimates $\Lambda_0$ by fitting the observed number of triggers to a
Poisson distribution.  We have chosen to leave $\Lambda_0$ as a free
parameter in our canonical analysis with a broad prior and infer it
from the observed data, rather than using the \pycbc{} background
estimate to constrain the prior, which would result in a much narrower
posterior on $\Lambda_0$.  This is equivalent to the \gstlal{}
procedure for $\Lambda_0$ estimation in the absence of signals; the
presence of a small number of signals in our data here do not
substantially change the $\Lambda_0$ estimate due to the overwhelming
number of background triggers in the data set.

\begin{figure}
  \plotone{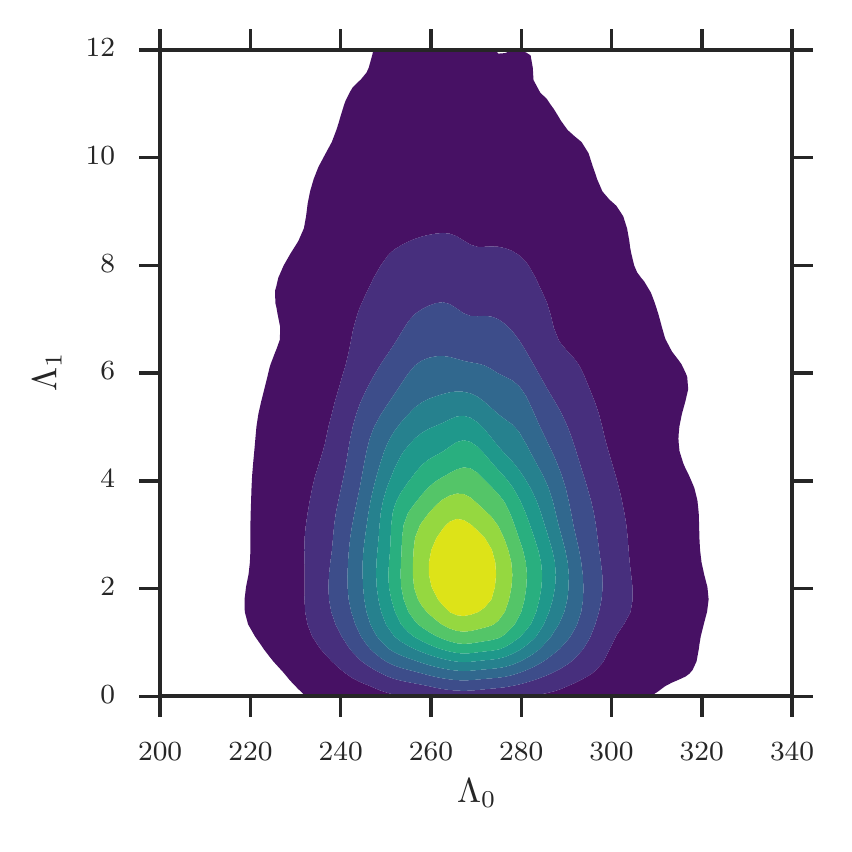}
  \caption{The two-dimensional posterior on terrestrial and
    astrophysical trigger expected counts ($\Lambda_0$ and $\Lambda_1$
    in Eq.\ \leteqtwopopposterior{} of the Letter) for the \pycbc{}
    search.  Contours are drawn at the 10\%, 20\%, \ldots, 90\%, and
    99\% credible levels.  There is no meaningful correlation between
    the two variables.  The Poisson uncertainty in the terrestrial
    count is $\sim \sqrt{270}$, or $16$, which is also very nearly the
    Poisson uncertainty in the total count.  Because this uncertainty
    is much larger than the astrophysical count, changes in the
    astrophysical count do not force the terrestrial count to adjust
    in a meaningful way and the variables are uncorrelated in the
    posterior.}
  \label{fig:Lambda0-Lambda1}
\end{figure}

Using a broad prior on $\Lambda_0$ is \emph{conservative} in the sense
that it will broaden the posterior on $\Lambda_1$ from which we infer
rates.  However, because there are so many more triggers in both
searches of terrestrial origin than astrophysical there is little
correlation between $\Lambda_0$ and $\Lambda_1$, and so there is
little difference between the posterior we obtain on $\Lambda_1$ and
the posterior we would have obtained had we implemented the tight
prior on $\Lambda_0$.  Figure \ref{fig:Lambda0-Lambda1} shows the
two-dimensional posterior we obtain from Eq.\ \leteqtwopopposterior{}
of the Letter on $\Lambda_0$ and $\Lambda_1$.

\begin{figure}
  \plotone{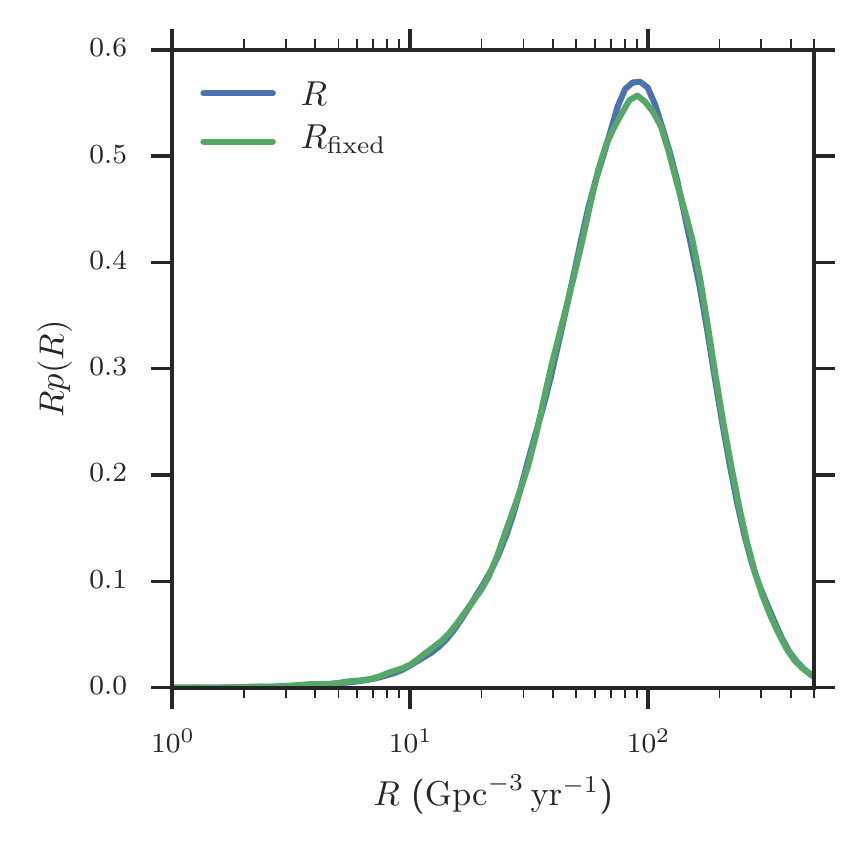}
  \caption{The posterior on the population-based rate obtained from
    our canonical analysis (blue) and an analysis where the expected
    background count, $\Lambda_0$, is fixed to the value measured by
    the \pycbc{} pipeline, $\Lambda_0 = \pycbcFixedBgCounts{}$
    (green).  There is no meaningful change in the rate posterior
    between the two analyses.}
  \label{fig:Lambda0-prior-sensitivity}
\end{figure}

We have checked that using a $\delta$-function prior
\begin{equation}
  \label{eq:pycbc-Lambda0-prior}
  p\left( \Lambda_0 \right) = \delta \left( \Lambda_0 - \pycbcFixedBgCounts{} \right)
\end{equation}
in the \pycbc{} analysis that is the result of the pipeline
$\Lambda_0$ estimate from timeslides\footnote{While the statistical
  uncertainty on the pipeline $\Lambda_0$ estimate is not precisely
  zero, $\sigma_{\Lambda_0}/ \Lambda_0 \lesssim 10^{-3}$, it is so
  small that a $\delta$-function prior is appropriate.}
\citep{GW150914-CBC} and using a looser prior that is the result of a
\gstlal{} estimate on a single set of time-slid data produces no
meaningful change in our results.  Figure
\ref{fig:Lambda0-prior-sensitivity} shows our canonical rate posterior
inferred with the \pycbc{} $\Lambda_0$ prior in Eq.\
\eqref{eq:pycbc-Lambda0-prior} and our canonical broad prior.

\section{Universal Astrophysical Trigger Distribution}
\label{suppsec:universal}

Both the \pycbc{} and \gstlal{} pipelines rely on the \ac{SNR} as part
of their detection statistic, $x$.  The \ac{SNR} of an astrophysical
trigger is a function of the detector noise at the time of detection
and the parameters of the trigger.  \citet{Schutz2011} and
\citet{Chen2014} demonstrate that the distribution of the expected
\ac{SNR} $\langle \rho \rangle$ in a simple model of a detection
pipeline that simply thresholds on \ac{SNR},
$\rho \geq \rho_\mathrm{th}$, with sources in the local universe is
\emph{universal}, that is, independent of the source properties.  It
follows
\begin{equation}
  \label{eq:snr-universal}
  p\left( \langle \rho \rangle \right) = \frac{3
    \rho_\mathrm{th}^3}{\langle \rho\rangle^4}.
\end{equation}
This result follows from the fact that the expected value of the
\ac{SNR} in a matched-filter search for \ac{CBC} signals scales
inversely with transverse comoving distance \citep{Hogg1999}:
\begin{equation}
  \label{eq:snr-scaling}
  \langle \rho \rangle = \frac{A\left( m_1, m_2, \vec{a}_1, \vec{a}_2, S(f), z \right) B \left(
      \mathrm{angles} \right)}{D_M},
\end{equation}
where $A$ is an amplitude factor that depends on the intrinsic
properties (source-frame masses and spins) of the source, the detector
sensitivity expressed as a noise power spectral density $S(f)$ as a
function of observer frequency and redshift $z$, and $B$ is an angular
factor depending on the location of the source in the sky and the
relative orientations of binary orbit and detector.  The redshift
enters $A$ only through shifting the source waveform to lower
frequency at higher redshift, changing $A$ because the sensitivity
varies with observer frequency $f$.  For the redshifts to which we are
sensitive to \ac{BBH} in this observation period this effect on $A$ is
small.

If we assume that the distribution of source parameters is constant
over the range of distances to which we are sensitive, and ignore the
small redshift-dependent sensitivity correction mentioned above, then
the distribution of \ac{SNR} will be governed entirely by the
distribution of distances of the sources, which, in the local universe
is approximately
\begin{equation}
  \label{eq:approx-dL-dist}
  p\left( D_M \right) \propto D_M^2,
\end{equation}
yielding the distribution of \ac{SNR} given in Eq.\
\eqref{eq:snr-universal}.  

\begin{figure}
  \plotone{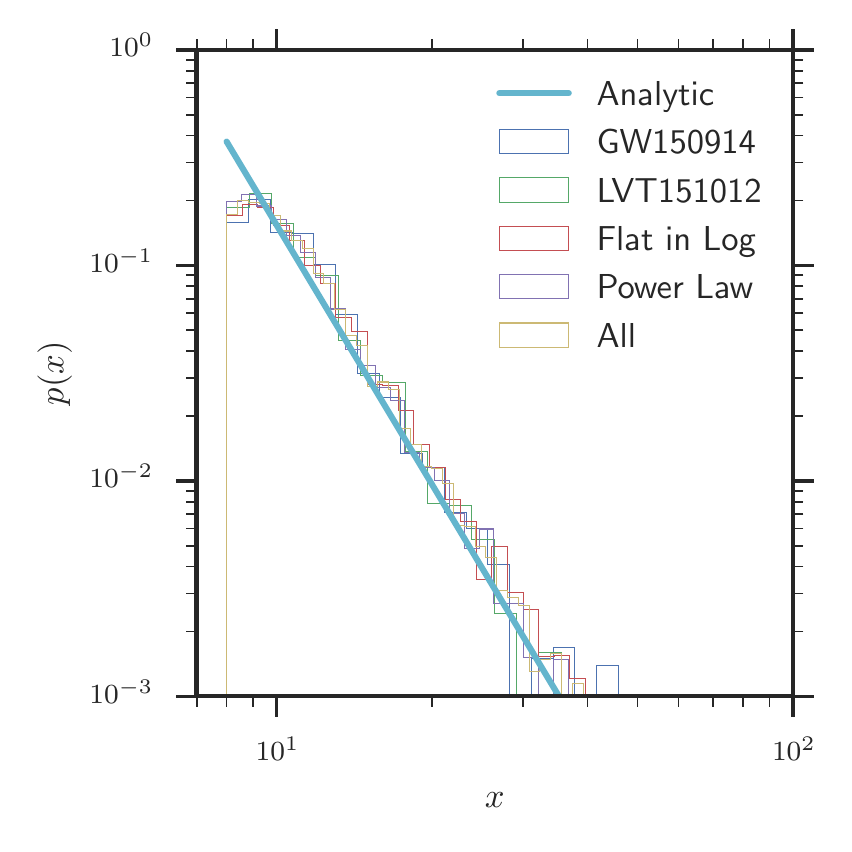}
  \caption{The distribution of detection statistics in the \pycbc{}
    pipeline for the signals recovered in the injection campaigns used
    to estimate sensitive time-volumes for various \ac{BBH} population
    assumptions (see Sections 2 and 3 of the Letter).  The solid line
    gives the analytic approximation to the distribution from Eq.\
    \eqref{eq:snr-universal}, which agrees well with the recovered
    statistics for loud signals; for quieter signals the interaction
    of various thresholds in the pipeline causes the distribution to
    deviate from the analytic approximation, but it remains
    independent of the source distribution.}
  \label{fig:universal}
\end{figure}

Both the \pycbc{} and \gstlal{} pipelines use goodness-of-fit
statistics in addition to \ac{SNR} and employ a more complicated
system of thresholds than this simple model, but the empirical
distribution of detection statistics remains, to an approximation
suitable for our purposes, independent of the source parameters.
Figure \ref{fig:universal} shows the distribution of recovered
detection statistics for the various injection campaigns with varying
source distribution used to estimate sensitive time-volumes in the
\pycbc{} pipeline.  In each injection campaign $\mathcal{O}(1000)$
signals were recovered.  For loud signals, the detection statistic is
proportional to \ac{SNR} in this pipeline, and the distribution is not
sensitive to the complicated thresholding in the pipeline, so we
recover Eq.\ \eqref{eq:snr-universal}; for quiet signals the
interaction of various single-detector thresolds in the pipeline
causes the distribution to deviate from this analytic approximation,
but it remains independent of the distribution of sources.  Note that
the \emph{empirical} distribution of detection statistics, not the
analytic one, forms the basis for $p_1$, the foreground distribution
used in this rate estimation work.

To quantify the deviations from universality, we have preformed
two-sample \ac{KS} tests between all six pairings of the sets of
detections statistics recovered in the injection campaigns described
in Sections 2 and 3 of the Letter and featured in Figure
\ref{fig:universal}.  The most extreme \ac{KS} $p$-value occurred with
the comparison between the injection set with \ac{BBH} masses drawn
flat in $\log m$ and the one with masses drawn from a power law (both
described in Section 3 of the Letter); this test gave a $p$-value of
$0.013$.  Given that we have performed six identical comparisons we
cannot reject the null hypothesis that the empirical distributions
used for rate estimation from the \pycbc{} pipeline are identical even
at the relatively weak significance $\alpha = 0.05$.  Certainly any
differences in detection statistic distribution attributable to the
\ac{BBH} population are far too small to matter with the few
astrophysical signals in our data set (compared with
$\mathcal{O}(1000)$ recovered injections in each campaign).

Because the distribution of detection statistics is, to a very good
approximation, \emph{universal}, we cannot learn anything about the
source population from the detection statistic alone; we must instead
resort to \ac{PE} followup \citep{Veitch2015,GW150914-PARAMESTIM} of
triggers to determine their parameters.  The parameters of the
waveform template that produced the trigger can be used to guess the
parameters of the source that generated that trigger, but the bias and
uncertainty in this estimate are very large compared to the \ac{PE}
estimate.  We therefore ignore the parameters of the waveform template
that generated the trigger in the assignment of triggers to \ac{BBH}
classes.

\section{Count Posterior}
\label{suppsec:count-posterior}

We impose a prior on the $\Lambda$ parameters of:
\begin{equation}
  \label{eq:two-pop-prior-supp}
  p\left(\Lambda_1,\Lambda_0\right) \propto
    \frac{1}{\sqrt{\Lambda_1}} \frac{1}{\sqrt{\Lambda_0}}.
\end{equation}

The posterior on expected counts is proportional to the product of the
likelihood from Eq.\ \leteqtwopoplikelihood{} of the Letter and the
prior from Eq.\ \eqref{eq:two-pop-prior-supp}:
\begin{multline}
  \label{eq:two-pop-posterior-supp}
  p\left(\Lambda_1, \Lambda_0 | \left\{ x_j | j = 1, \ldots, M\right\}
  \right) \\ \propto \left\{ \prod_{j = 1}^{M} \left[ \Lambda_1
    p_1\left(x_j\right) + \Lambda_0 p_0\left( x_j \right) \right] \right\}
  \\ \times \exp\left[ -\Lambda_1 - \Lambda_0 \right]
  \frac{1}{\sqrt{\Lambda_1 \Lambda_0}}.
\end{multline}
For estimation of the Poisson rate parameter in a simple Poisson
model, the Jeffreys prior is $1/\sqrt{\Lambda}$.  With this prior, the
posterior mean on $\Lambda$ is $N+1/2$ for $N$ observed counts.  With
a prior proportional to $1/\Lambda$ the mean is $N$ for $N>0$, but the
posterior is improper when $N=0$.  For a flat prior, the mean is
$N+1$.  Though the behaviour of the mean is not identical with our
mixture model posterior, it is similar; because we find
$\left\langle \Lambda_1 \right\rangle \gg 1/2$, the choice of prior
among these three reasonable options has little influence on our
results here.

For the \pycbc{} data set we find the posterior median and $90\%$
credible range $\Lambda_1 = \countonetwopop$ above our threshold.  For
the \gstlal{} set we find the posterior median and 90\% credible range
$\Lambda_1=\countonetwopopGSTLAL{}$.  Though we have only one event
(\firstevent{}) at exceptionally high significance, and one other at
marginal significance (\secondevent{}), the counting analysis shows
these to be consistent with the possible presence of several more
events of astrophysical origin at lower detection statistic in both
pipelines.

The thresholds applied to the \pycbc{} and \gstlal{} triggers for this
analysis are \emph{not} equivalent to each other in terms of either
\ac{SNR} or false alarm rate; instead, both thresholds have been
chosen so that the rate of triggers of terrestrial origin
($\Lambda_0 p_0$) dominates near threshold.  
Since the threshold is set at \emph{different} values for each
pipeline, we do not expect the counts to be the same between pipelines.

\begin{figure*}
  \plotone{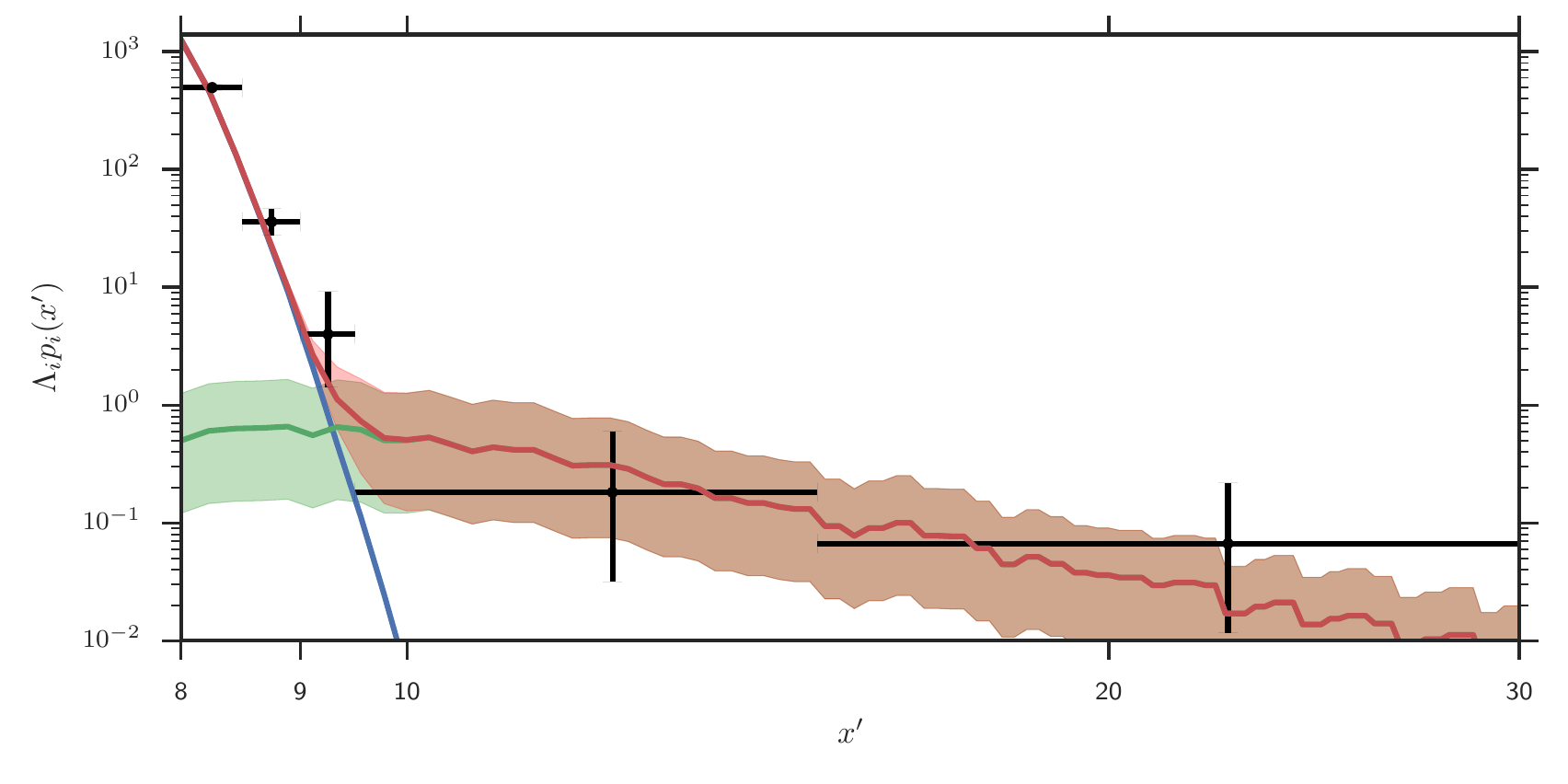}
  \caption{The inferred number density of astrophysical (green),
    terrestrial (blue), and all (red) triggers as a function of $x'$
    for the \pycbc{} search (cf.\ Eq.~\leteqtwopopfgmcrate{} of the Letter),
    using the models for each population described in Section
    \letseccounts{} of the Letter.  The solid lines give the posterior median and
    the shaded regions give the symmetric 90\% credible interval from
    the posterior in Eq.\ \leteqtwopopposterior{} of the Letter. We also show a
    binned estimate of the trigger number density from the search
    (black); bars indicate the 68\% confidence Poisson uncertainty on
    the number of triggers in the vertical-direction and bin width in
    the horizontal-direction.}
  \label{fig:foreground-background}
\end{figure*}

The estimated astrophysical and terrestrial trigger rate densities
(Eq.\ \leteqtwopopfgmcrate{} of the Letter) for \pycbc{} are plotted in
Figure~\ref{fig:foreground-background}.  We select triggers from a
subset of the search parameter space (i.e.\ our bank of template
waveforms) that contains \firstevent{} as well as the mass range
considered for possible alternative populations of \ac{BBH} binaries
in Section \letsecmassdistribution{} of the Letter.  There are $M' = \mcoincs{}$
two-detector coincident triggers in this range in the \pycbc{} search
\citep{GW150914-CBC}.  Figure \ref{fig:foreground-background} also
shows an estimate of the density of triggers that comprise our data
set which agrees well with our inference of the trigger rate.

\begin{figure*}
  \plottwo{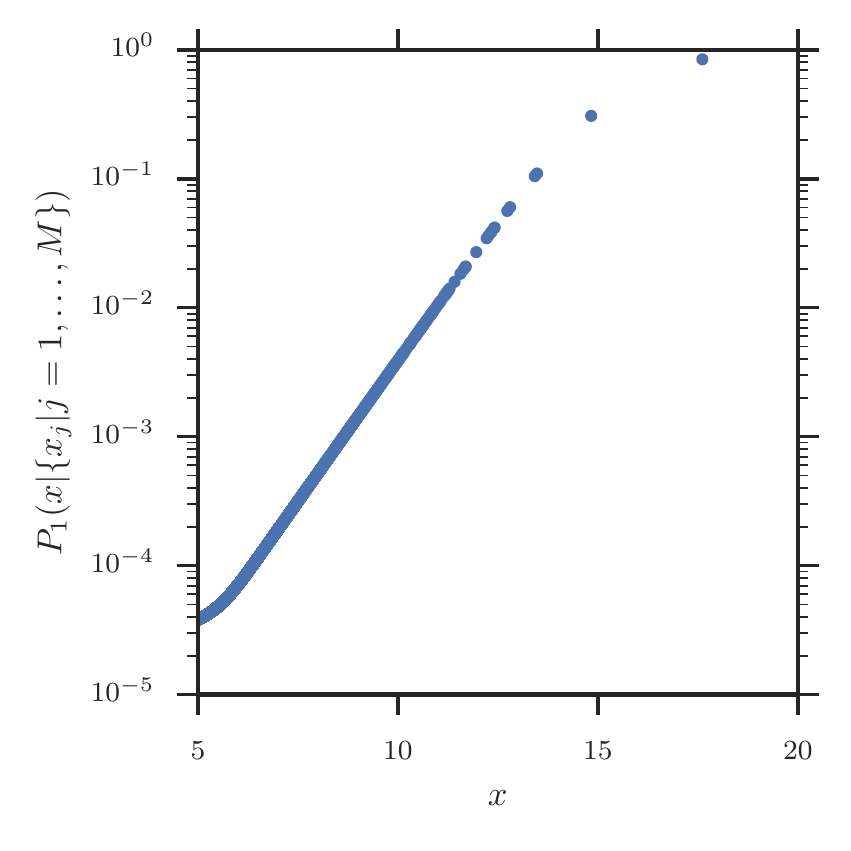}{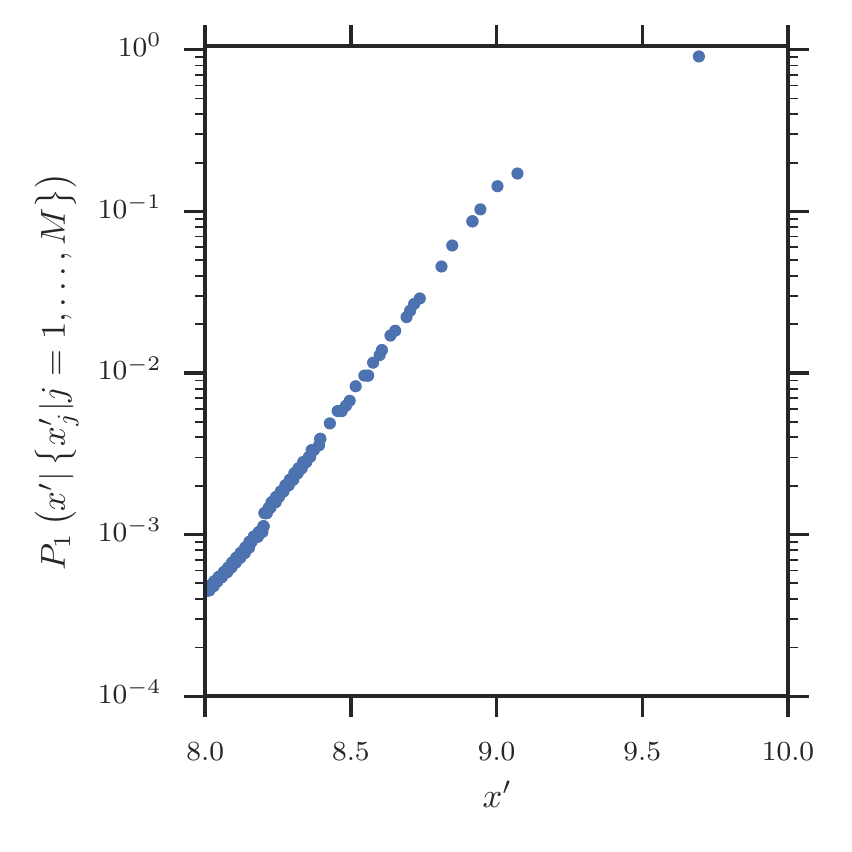}
  \caption{The posterior probability that coincident triggers in our
    analysis come from an astrophysical source (see Eq.~\leteqpfore{}
    of the Letter), taking into account the astrophysical and
    terrestrial expected counts estimated in Section \letseccounts{}
    of the Letter.  Left: the \gstlal{} triggers with $x>5$; right:
    \pycbc{} triggers with $x'>8$.  \firstevent{} is not shown in the
    plot because its probability of astrophysical origin is
    effectively 100\%.  The only two triggers with $P_1 \gtrsim 50\%$
    are \firstevent{} and \secondevent{}.  For \firstevent{}, we find
    $P_1 = 1$ to very high precision; for \secondevent{}, the
    \gstlal{} pipeline finds $P_1 = \secondpforeGSTLAL{}$ and the
    \pycbc{} pipeline finds $P_1 = \secondpfore{}$.}
  \label{fig:pfore}
\end{figure*}

Based on the probability of astrophysical origin inferred for
\secondevent{} from the two-component mixture model in Eq.\
\eqref{eq:two-pop-posterior-supp} and shown in Figure \ref{fig:pfore},
we introduce a third class of signals and use a three-component
mixture model with expected counts $\Lambda_0$ (terrestrial),
$\Lambda_1$ (\firstevent{}-like), and $\Lambda_2$
(\secondevent{}-like) to infer rates in Sections \letseccounts{} of
the Letter and \letsecrates{} of the Letter.

We use the Stan and \emcee{} Markov-Chain Monte Carlo samplers
\citep{Foreman-Mackey2013,stan2015,pystan2015} to draw samples from
the posterior in Eq.\ \leteqtwopopposterior{} of the Letter for the two
pipelines.  We have assessed the convergence and mixing of our chains
using empirical estimates of the autocorrelation length in each
parameter \citep{Sokal1996}, the Gelman-Rubin $R$ convergence
statistic \citep{Gelman1992}, and through visual inspection of chain
plots.  By all measures, the chains appear well-converged to the
posterior distribution. 

Table \ref{tab:count-vt-table} contains the full results on expected
counts and associated sensitive time-volumes for both pipelines.

\begin{deluxetable*}{lcccc}
  \tablewidth{0pt}
  \tablecolumns{5} \tablecaption{Expected counts and sensitive
    time-volumes to \ac{BBH} mergers estimated under various
    assumptions.  See Sections \letseccounts{} of the Letter,
    \letsecrates{} of the Letter, \letsecmassdistribution{} of the
    Letter and \ref{suppsec:count-posterior}.
    \label{tab:count-vt-table}}
  \tablehead{  & \multicolumn{2}{c}{$\Lambda$} &
    \multicolumn{2}{c}{$\avgVT / \gpcyr$} \\
    & \pycbc{} & \gstlal{} & \pycbc{} & \gstlal{} }
  \startdata
  \firstevent{} & \countone{} & \countoneGSTLAL{} & \sensVTone{} & \sensVToneGSTLALnoUnits{} \\
  \secondevent{} & \counttwo{} & \counttwoGSTLAL{} & \sensVTtwo{} & \sensVTtwoGSTLALnoUnits{} \\
  Both & \countall{} & \countallGSTLAL{} & \nodata & \nodata \\
  \cutinhead{Astrophysical}
  Flat in log mass & \multirow{ 2}{*}{\countonetwopop} & \multirow{ 2}{*}{\countonetwopopGSTLAL} & \sensVTflatlog{} & \sensVTflatlogGSTLALnoUnits{} \\
  Power Law (-2.35) & & & \sensVTpowerlaw{} & \sensVTpowerlawGSTLALnoUnits{} \\
  \enddata
\end{deluxetable*}

\section{Calibration Uncertainty}
\label{suppsec:calibration-uncert}

The LIGO detectors are subject to uncertainty in their calibration, in
both the measured amplitude and phase of the gravitational-wave
strain.  \citet{GW150914-CALIBRATION} discusses the methods used to
calibrate the strain output of the detector during the \OBSDAYS{} of
coincident observations discussed here.  \citet{GW150914-CALIBRATION}
estimates that the reported strain is accurate to within 10\% in
amplitude and 10 degrees in phase between $20\,\mathrm{Hz}$ and
$1\,\mathrm{kHz}$ throughout the observations.  

The \acp{SNR} reported by our searches are quadratically sensitive to
calibration errors because they are maximized over arrival time,
waveform phase, and a template bank of waveforms
\citep{Allen1996,Brown2004}.  \citet{GW150914-CBC} demonstrates that
the other search pipeline outputs are also not affected to a
significant degree by the calibration uncertainty present during our
observing run.  Therefore, we ignore effects of calibration on the
pipeline detection statistics $x$ and $x'$ we use here to estimate
rates from the \pycbc{} and \gstlal{} pipelines.

The amplitude calibration uncertainty in the detector results at
leading order in a corresponding uncertainty between the luminosity
distances of sources measured from real detector outputs
\citep{GW150914-PARAMESTIM} and the luminosity distances used to
produce injected waveforms used to estimate sensitive time-volumes in
this work.  A 10\% uncertainty in $d_L$ at these redshifts corresponds
to an approximately 30\% uncertainty in volume.  We model this
uncertainty by treating $\avgVT$ as a parameter in our analysis, and
imposing a log-normal prior:
\begin{equation}
  \label{eq:VT-prior}
  p\left( \log \avgVT \right) \propto N\left( \log \mu,
    \frac{\sigma}{\mu} \right),
\end{equation}
where $\mu$ is the Monte-Carlo estimate of sensitive time-volume
produced from the injection campaigns described in Section
\letsecrates{} of the Letter and
\begin{equation}
  \label{eq:VT-sigma}
  \sigma^2 = \sigma_\mathrm{cal}^2 + \sigma_\mathrm{stat}^2,
\end{equation}
with $\sigma_\mathrm{cal} = 0.3 \mu$ and $\sigma_\mathrm{stat}$ is the
estimate of the Monte-Carlo uncertainty from the finite number of
recovered injections reported above.  In all cases
$\sigma_\mathrm{cal} \gg \sigma_\mathrm{stat}$.  

Since the likelihood in Eqs.\ \leteqtwopoplikelihood{} of the Letter or
\leteqlikelihood{} of the Letter does not constrain $\avgVT$ independently of
$R$, sampling over $\avgVT$ at the same time as $\Lambda$ and $R$ has
the effect of convolving the log-normal distribution of $\avgVT$ with
the posterior on $\Lambda$ in the inference of $R$.  In spite of the
30\% relative uncertainty in $\avgVT$ from calibration uncertainty,
the counting uncertainty on $R$ from the small number of detected
events dominates the width of the posterior on $R$.

\section{Analytic Sensitivity Estimate}
\label{suppsec:analytic-vt}
\begin{figure}
  \plotone{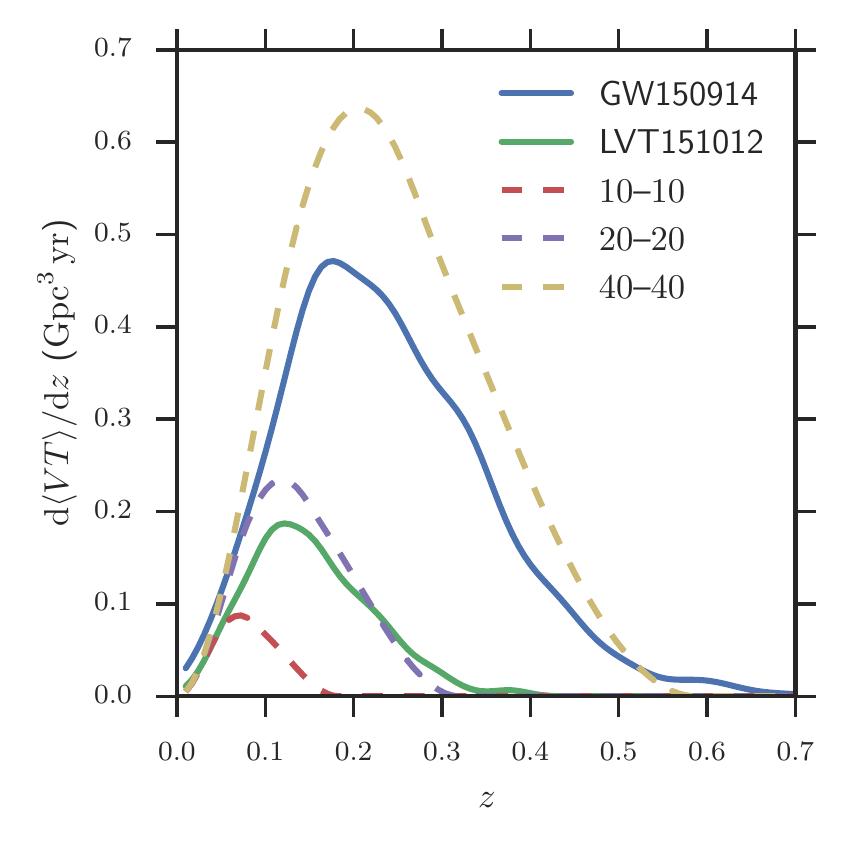}
  \caption{The rate at which sensitive time-volume accumulates with
    redshift.  Curves labeled by component masses in $\Msun$ are
    computed using the approximate prescription described in Section
    \ref{suppsec:analytic-vt}, assuming sources with fixed masses in
    the comoving frame and without spin; the \firstevent{} and
    \secondevent{} curves are determined from the Monte-Carlo
    injection campaign described in Section \letsecrates{} of the
    Letter.}
  \label{fig:dVTdz}
\end{figure}

As a rough check on our $\avgVT{}$ estimates and the integrand
$\mathrm{d} \avgVT / \mathrm{d}z$, we find that the following
approximate, analytic procedure also produces a good approximation to
the pycbc Monte-Carlo estimate in Table \ref{tab:count-vt-table}.
\begin{enumerate}
\item Generate inspiral--merger--ringdown waveforms in a single detector
  at various redshifts from the source distribution $s(\theta)$ with
  random orientations and sky positions.
\item Using the high-sensitivity early Advanced LIGO noise power
  spectral density from \citet{Aasi2013ObsScenario}, compute the
  \ac{SNR} in a single detector.
\item Consider a signal found if the \ac{SNR} is greater than
  \approxVTthresh{}.
\end{enumerate}
Employed with the source distributions described above, this
approximate procedure yields $\avgVT_1 \simeq \approxVTone$ and
$\avgVT_2 \simeq \approxVTtwo$ for the sensitivity to the two classes
of merging \ac{BBH} system.  Figure \ref{fig:dVTdz} shows the
sensitive time-volume integrand,
\begin{equation}
  \label{eq:vt-integrand}
  \diff{\avgVT{}}{z} \equiv T \frac{1}{1+z} \diff{V_c}{z} \int \dd
  \theta \, s(\theta) f(z, \theta)
\end{equation}
estimated from this procedure for systems with various parameters
superimposed on the Monte-Carlo estimates from the injection campaign
described above.

\acknowledgments

\input{LVCacknowledgments.tex}

\bibliographystyle{aasjournal}
\bibliography{LIGO-P1500217_GW150914_Rates,../macros/GW150914_refs}

\allauthors

\end{document}

%% file: macros.tex
\newcommand{\macro}[1]{\textcolor{red}{#1}} 

\newcommand\SECONDMONDAY{\macro{LVT151012}}

\newcommand{\Msun}{\ensuremath{\mathrm{M}_\odot}}

\newcommand{\OBSDAYS}{\macro{\ensuremath{\macro{16}~\mathrm{days}}}}

\newcommand{\ninetyrange}[3]{\ensuremath{{#1}^{+{#2}}_{-{#3}}}}
\newcommand{\ninetyinterval}[2]{\ensuremath{{#1}\text{--}{#2}}}

\newcommand{\gpcyr}{\ensuremath{\mathrm{Gpc}^3\,\mathrm{yr}}}
\newcommand{\pergpcyr}{\ensuremath{\mathrm{Gpc}^{-3}\,\mathrm{yr}^{-1}}}

\newcommand{\oneraterangetorulethemall}{\macro{\ensuremath{\ninetyinterval{2}{600} \, \pergpcyr}}}

\newcommand{\secondpforeGSTLAL}{\macro{\ensuremath{0.84}}}
\newcommand{\secondpfore}{\macro{\ensuremath{0.91}}}

\newcommand{\newsnrthresh}{\macro{\ensuremath{8}}}
\newcommand{\loudnewsnr}{\macro{\ensuremath{10.1}}}

\newcommand{\pycbcFixedBgCounts}{\macro{270}}

\newcommand{\countonetwopop}{\macro{\ensuremath{\ninetyrange{3.2}{4.9}{2.4}}}}
\newcommand{\countonetwopopGSTLAL}{\macro{\ensuremath{\ninetyrange{4.8}{7.9}{3.8}}}}

\newcommand{\mcoincs}{\macro{\ensuremath{270}}}

\newcommand{\countone}{\macro{\ensuremath{\ninetyrange{2.1}{4.1}{1.7}}}}
\newcommand{\counttwo}{\macro{\ensuremath{\ninetyrange{2.0}{4.0}{1.7}}}}
\newcommand{\countoneGSTLAL}{\macro{\ensuremath{\ninetyrange{3.6}{6.9}{2.9}}}}
\newcommand{\counttwoGSTLAL}{\macro{\ensuremath{\ninetyrange{3.0}{6.8}{2.7}}}}
\newcommand{\countall}{\macro{\ensuremath{\ninetyrange{4.5}{5.5}{3.1}}}}
\newcommand{\countallGSTLAL}{\macro{\ensuremath{\ninetyrange{7.4}{9.2}{5.1}}}}

\newcommand{\sensVTone}{\macro{\ensuremath{\ninetyrange{0.130}{0.084}{0.051}}}}
\newcommand{\sensVTtwo}{\macro{\ensuremath{\ninetyrange{0.032}{0.020}{0.012}}}}
\newcommand{\sensVTflatlog}{\macro{\ensuremath{\ninetyrange{0.050}{0.032}{0.019}}}}
\newcommand{\sensVTpowerlaw}{\macro{\ensuremath{\ninetyrange{0.0154}{0.0098}{0.0060}}}}
\newcommand{\sensVToneGSTLALnoUnits}{\macro{\ensuremath{\ninetyrange{0.21}{0.14}{0.08}}}}
\newcommand{\sensVTtwoGSTLALnoUnits}{\macro{\ensuremath{\ninetyrange{0.048}{0.031}{0.019}}}}
\newcommand{\sensVTflatlogGSTLALnoUnits}{\macro{\ensuremath{\ninetyrange{0.080}{0.051}{0.031}}}}
\newcommand{\sensVTpowerlawGSTLALnoUnits}{\macro{\ensuremath{\ninetyrange{0.024}{0.015}{0.009}}}}

\newcommand{\approxVTthresh}{\macro{\ensuremath{8}}}

\newcommand{\approxVTone}{\macro{\ensuremath{0.107\,\gpcyr}}}
\newcommand{\approxVTtwo}{\macro{\ensuremath{0.0225\,\gpcyr}}}

%% file: authors.tex
\author{%
B.~P.~Abbott,\altaffilmark{1}}  
\author{
R.~Abbott,\altaffilmark{1}  
T.~D.~Abbott,\altaffilmark{2}  
M.~R.~Abernathy,\altaffilmark{1}  
F.~Acernese,\altaffilmark{3,4}
K.~Ackley,\altaffilmark{5}  
C.~Adams,\altaffilmark{6}  
T.~Adams,\altaffilmark{7}
P.~Addesso,\altaffilmark{3}  
R.~X.~Adhikari,\altaffilmark{1}
% ...
}  

\medskip
\affiliation {$^{\dag}$Deceased, May 2015. $^{\ddag}$Deceased, March 2015.
\\
{(LIGO Scientific Collaboration and Virgo Collaboration)}%
}%
\medskip

\altaffiltext {1}{LIGO, California Institute of Technology, Pasadena, CA 91125, USA }
\altaffiltext {2}{Louisiana State University, Baton Rouge, LA 70803, USA }
\altaffiltext {3}{Universit\`a di Salerno, Fisciano, I-84084 Salerno, Italy }
\altaffiltext {4}{INFN, Sezione di Napoli, Complesso Universitario di Monte S.Angelo, I-80126 Napoli, Italy }
\altaffiltext {5}{University of Florida, Gainesville, FL 32611, USA }
\altaffiltext {6}{LIGO Livingston Observatory, Livingston, LA 70754, USA }
\altaffiltext {7}{Laboratoire d'Annecy-le-Vieux de Physique des Particules (LAPP), Universit\'e Savoie Mont Blanc, CNRS/IN2P3, F-74941 Annecy-le-Vieux, France }
\altaffiltext {8}{Albert-Einstein-Institut, Max-Planck-Institut f\"ur Gravi\-ta\-tions\-physik, D-30167 Hannover, Germany }
\altaffiltext {9}{Nikhef, Science Park, 1098 XG Amsterdam, Netherlands }
\altaffiltext {10}{LIGO, Massachusetts Institute of Technology, Cambridge, MA 02139, USA }
\altaffiltext {11}{Instituto Nacional de Pesquisas Espaciais, 12227-010 S\~{a}o Jos\'{e} dos Campos, S\~{a}o Paulo, Brazil }
\altaffiltext {12}{INFN, Gran Sasso Science Institute, I-67100 L'Aquila, Italy }
\altaffiltext {13}{INFN, Sezione di Roma Tor Vergata, I-00133 Roma, Italy }
\altaffiltext {14}{Inter-University Centre for Astronomy and Astrophysics, Pune 411007, India }
\altaffiltext {15}{International Centre for Theoretical Sciences, Tata Institute of Fundamental Research, Bangalore 560012, India }
\altaffiltext {16}{University of Wisconsin-Milwaukee, Milwaukee, WI 53201, USA }
\altaffiltext {17}{Leibniz Universit\"at Hannover, D-30167 Hannover, Germany }
\altaffiltext {18}{Universit\`a di Pisa, I-56127 Pisa, Italy }
\altaffiltext {19}{INFN, Sezione di Pisa, I-56127 Pisa, Italy }
\altaffiltext {20}{Australian National University, Canberra, Australian Capital Territory 0200, Australia }
\altaffiltext {21}{The University of Mississippi, University, MS 38677, USA }
\altaffiltext {22}{California State University Fullerton, Fullerton, CA 92831, USA }
\altaffiltext {23}{LAL, Universit\'e Paris-Sud, CNRS/IN2P3, Universit\'e Paris-Saclay, 91400 Orsay, France }
\altaffiltext {24}{Chennai Mathematical Institute, Chennai 603103, India }
\altaffiltext {25}{Universit\`a di Roma Tor Vergata, I-00133 Roma, Italy }
\altaffiltext {26}{University of Southampton, Southampton SO17 1BJ, United Kingdom }
\altaffiltext {27}{Universit\"at Hamburg, D-22761 Hamburg, Germany }
\altaffiltext {28}{INFN, Sezione di Roma, I-00185 Roma, Italy }
\altaffiltext {29}{Albert-Einstein-Institut, Max-Planck-Institut f\"ur Gravitations\-physik, D-14476 Potsdam-Golm, Germany }
\altaffiltext {30}{APC, AstroParticule et Cosmologie, Universit\'e Paris Diderot, CNRS/IN2P3, CEA/Irfu, Observatoire de Paris, Sorbonne Paris Cit\'e, F-75205 Paris Cedex 13, France }
\altaffiltext {31}{Montana State University, Bozeman, MT 59717, USA }
\altaffiltext {32}{Universit\`a di Perugia, I-06123 Perugia, Italy }
\altaffiltext {33}{INFN, Sezione di Perugia, I-06123 Perugia, Italy }
\altaffiltext {34}{European Gravitational Observatory (EGO), I-56021 Cascina, Pisa, Italy }
\altaffiltext {35}{Syracuse University, Syracuse, NY 13244, USA }
\altaffiltext {36}{SUPA, University of Glasgow, Glasgow G12 8QQ, United Kingdom }
\altaffiltext {37}{LIGO Hanford Observatory, Richland, WA 99352, USA }
\altaffiltext {38}{Wigner RCP, RMKI, H-1121 Budapest, Konkoly Thege Mikl\'os \'ut 29-33, Hungary }
\altaffiltext {39}{Columbia University, New York, NY 10027, USA }
\altaffiltext {40}{Stanford University, Stanford, CA 94305, USA }
\altaffiltext {41}{Universit\`a di Padova, Dipartimento di Fisica e Astronomia, I-35131 Padova, Italy }
\altaffiltext {42}{INFN, Sezione di Padova, I-35131 Padova, Italy }
\altaffiltext {43}{CAMK-PAN, 00-716 Warsaw, Poland }
\altaffiltext {44}{University of Birmingham, Birmingham B15 2TT, United Kingdom }
\altaffiltext {45}{Universit\`a degli Studi di Genova, I-16146 Genova, Italy }
\altaffiltext {46}{INFN, Sezione di Genova, I-16146 Genova, Italy }
\altaffiltext {47}{RRCAT, Indore MP 452013, India }
\altaffiltext {48}{Faculty of Physics, Lomonosov Moscow State University, Moscow 119991, Russia }
\altaffiltext {49}{SUPA, University of the West of Scotland, Paisley PA1 2BE, United Kingdom }
\altaffiltext {50}{University of Western Australia, Crawley, Western Australia 6009, Australia }
\altaffiltext {51}{Department of Astrophysics/IMAPP, Radboud University Nijmegen, P.O. Box 9010, 6500 GL Nijmegen, Netherlands }
\altaffiltext {52}{Artemis, Universit\'e C\^ote d'Azur, CNRS, Observatoire C\^ote d'Azur, CS 34229, Nice cedex 4, France }
\altaffiltext {53}{MTA E\"otv\"os University, ``Lendulet'' Astrophysics Research Group, Budapest 1117, Hungary }
\altaffiltext {54}{Institut de Physique de Rennes, CNRS, Universit\'e de Rennes 1, F-35042 Rennes, France }
\altaffiltext {55}{Washington State University, Pullman, WA 99164, USA }
\altaffiltext {56}{Universit\`a degli Studi di Urbino ``Carlo Bo,'' I-61029 Urbino, Italy }
\altaffiltext {57}{INFN, Sezione di Firenze, I-50019 Sesto Fiorentino, Firenze, Italy }
\altaffiltext {58}{University of Oregon, Eugene, OR 97403, USA }
\altaffiltext {59}{Laboratoire Kastler Brossel, UPMC-Sorbonne Universit\'es, CNRS, ENS-PSL Research University, Coll\`ege de France, F-75005 Paris, France }
\altaffiltext {60}{Astronomical Observatory Warsaw University, 00-478 Warsaw, Poland }
\altaffiltext {61}{VU University Amsterdam, 1081 HV Amsterdam, Netherlands }
\altaffiltext {62}{University of Maryland, College Park, MD 20742, USA }
\altaffiltext {63}{Center for Relativistic Astrophysics and School of Physics, Georgia Institute of Technology, Atlanta, GA 30332, USA }
\altaffiltext {64}{Institut Lumi\`{e}re Mati\`{e}re, Universit\'{e} de Lyon, Universit\'{e} Claude Bernard Lyon 1, UMR CNRS 5306, 69622 Villeurbanne, France }
\altaffiltext {65}{Laboratoire des Mat\'eriaux Avanc\'es (LMA), IN2P3/CNRS, Universit\'e de Lyon, F-69622 Villeurbanne, Lyon, France }
\altaffiltext {66}{Universitat de les Illes Balears, IAC3---IEEC, E-07122 Palma de Mallorca, Spain }
\altaffiltext {67}{Universit\`a di Napoli ``Federico II,'' Complesso Universitario di Monte S.Angelo, I-80126 Napoli, Italy }
\altaffiltext {68}{NASA/Goddard Space Flight Center, Greenbelt, MD 20771, USA }
\altaffiltext {69}{Canadian Institute for Theoretical Astrophysics, University of Toronto, Toronto, Ontario M5S 3H8, Canada }
\altaffiltext {70}{Tsinghua University, Beijing 100084, China }
\altaffiltext {71}{Texas Tech University, Lubbock, TX 79409, USA }
\altaffiltext {72}{The Pennsylvania State University, University Park, PA 16802, USA }
\altaffiltext {73}{National Tsing Hua University, Hsinchu City, 30013 Taiwan, Republic of China }
\altaffiltext {74}{Charles Sturt University, Wagga Wagga, New South Wales 2678, Australia }
\altaffiltext {75}{University of Chicago, Chicago, IL 60637, USA }
\altaffiltext {76}{Caltech CaRT, Pasadena, CA 91125, USA }
\altaffiltext {77}{Korea Institute of Science and Technology Information, Daejeon 305-806, Korea }
\altaffiltext {78}{Carleton College, Northfield, MN 55057, USA }
\altaffiltext {79}{Universit\`a di Roma ``La Sapienza,'' I-00185 Roma, Italy }
\altaffiltext {80}{University of Brussels, Brussels 1050, Belgium }
\altaffiltext {81}{Sonoma State University, Rohnert Park, CA 94928, USA }
\altaffiltext {82}{Northwestern University, Evanston, IL 60208, USA }
\altaffiltext {83}{University of Minnesota, Minneapolis, MN 55455, USA }
\altaffiltext {84}{The University of Melbourne, Parkville, Victoria 3010, Australia }
\altaffiltext {85}{The University of Texas Rio Grande Valley, Brownsville, TX 78520, USA }
\altaffiltext {86}{The University of Sheffield, Sheffield S10 2TN, United Kingdom }
\altaffiltext {87}{University of Sannio at Benevento, I-82100 Benevento, Italy and INFN, Sezione di Napoli, I-80100 Napoli, Italy }
\altaffiltext {88}{Montclair State University, Montclair, NJ 07043, USA }
\altaffiltext {89}{Universit\`a di Trento, Dipartimento di Fisica, I-38123 Povo, Trento, Italy }
\altaffiltext {90}{INFN, Trento Institute for Fundamental Physics and Applications, I-38123 Povo, Trento, Italy }
\altaffiltext {91}{Cardiff University, Cardiff CF24 3AA, United Kingdom }
\altaffiltext {92}{National Astronomical Observatory of Japan, 2-21-1 Osawa, Mitaka, Tokyo 181-8588, Japan }
\altaffiltext {93}{School of Mathematics, University of Edinburgh, Edinburgh EH9 3FD, United Kingdom }
\altaffiltext {94}{Indian Institute of Technology, Gandhinagar Ahmedabad Gujarat 382424, India }
\altaffiltext {95}{Institute for Plasma Research, Bhat, Gandhinagar 382428, India }
\altaffiltext {96}{University of Szeged, D\'om t\'er 9, Szeged 6720, Hungary }
\altaffiltext {97}{Embry-Riddle Aeronautical University, Prescott, AZ 86301, USA }
\altaffiltext {98}{University of Michigan, Ann Arbor, MI 48109, USA }
\altaffiltext {99}{Tata Institute of Fundamental Research, Mumbai 400005, India }
\altaffiltext {100}{American University, Washington, D.C. 20016, USA }
\altaffiltext {101}{University of Massachusetts-Amherst, Amherst, MA 01003, USA }
\altaffiltext {102}{University of Adelaide, Adelaide, South Australia 5005, Australia }
\altaffiltext {103}{West Virginia University, Morgantown, WV 26506, USA }
\altaffiltext {104}{University of Bia{\l }ystok, 15-424 Bia{\l }ystok, Poland }
\altaffiltext {105}{SUPA, University of Strathclyde, Glasgow G1 1XQ, United Kingdom }
\altaffiltext {106}{IISER-TVM, CET Campus, Trivandrum Kerala 695016, India }
\altaffiltext {107}{Institute of Applied Physics, Nizhny Novgorod, 603950, Russia }
\altaffiltext {108}{Pusan National University, Busan 609-735, Korea }
\altaffiltext {109}{Hanyang University, Seoul 133-791, Korea }
\altaffiltext {110}{NCBJ, 05-400 \'Swierk-Otwock, Poland }
\altaffiltext {111}{IM-PAN, 00-956 Warsaw, Poland }
\altaffiltext {112}{Rochester Institute of Technology, Rochester, NY 14623, USA }
\altaffiltext {113}{Monash University, Victoria 3800, Australia }
\altaffiltext {114}{Seoul National University, Seoul 151-742, Korea }
\altaffiltext {115}{University of Alabama in Huntsville, Huntsville, AL 35899, USA }
\altaffiltext {116}{ESPCI, CNRS, F-75005 Paris, France }
\altaffiltext {117}{Universit\`a di Camerino, Dipartimento di Fisica, I-62032 Camerino, Italy }
\altaffiltext {118}{Southern University and A\&M College, Baton Rouge, LA 70813, USA }
\altaffiltext {119}{College of William and Mary, Williamsburg, VA 23187, USA }
\altaffiltext {120}{Instituto de F\'\i sica Te\'orica, University Estadual Paulista/ICTP South American Institute for Fundamental Research, S\~ao Paulo SP 01140-070, Brazil }
\altaffiltext {121}{University of Cambridge, Cambridge CB2 1TN, United Kingdom }
\altaffiltext {122}{IISER-Kolkata, Mohanpur, West Bengal 741252, India }
\altaffiltext {123}{Rutherford Appleton Laboratory, HSIC, Chilton, Didcot, Oxon OX11 0QX, United Kingdom }
\altaffiltext {124}{Whitman College, 345 Boyer Avenue, Walla Walla, WA 99362 USA }
\altaffiltext {125}{National Institute for Mathematical Sciences, Daejeon 305-390, Korea }
\altaffiltext {126}{Hobart and William Smith Colleges, Geneva, NY 14456, USA }
\altaffiltext {127}{Janusz Gil Institute of Astronomy, University of Zielona G\'ora, 65-265 Zielona G\'ora, Poland }
\altaffiltext {128}{Andrews University, Berrien Springs, MI 49104, USA }
\altaffiltext {129}{Universit\`a di Siena, I-53100 Siena, Italy }
\altaffiltext {130}{Trinity University, San Antonio, TX 78212, USA }
\altaffiltext {131}{University of Washington, Seattle, WA 98195, USA }
\altaffiltext {132}{Kenyon College, Gambier, OH 43022, USA }
\altaffiltext {133}{Abilene Christian University, Abilene, TX 79699, USA }

%% file: LVCacknowledgments.tex
The authors gratefully acknowledge the support of the United States
National Science Foundation (NSF) for the construction and operation of the
LIGO Laboratory and Advanced LIGO as well as the Science and Technology Facilities Council (STFC) of the
United Kingdom, the Max-Planck-Society (MPS), and the State of
Niedersachsen/Germany for support of the construction of Advanced LIGO 
and construction and operation of the GEO600 detector. 
Additional support for Advanced LIGO was provided by the Australian Research Council.
The authors gratefully acknowledge the Italian Istituto Nazionale di Fisica Nucleare (INFN),  
the French Centre National de la Recherche Scientifique (CNRS) and
the Foundation for Fundamental Research on Matter supported by the Netherlands Organisation for Scientific Research, 
for the construction and operation of the Virgo detector
and the creation and support  of the EGO consortium. 
The authors also gratefully acknowledge research support from these agencies as well as by 
the Council of Scientific and Industrial Research of India, 
Department of Science and Technology, India,
Science \& Engineering Research Board (SERB), India,
Ministry of Human Resource Development, India,
the Spanish Ministerio de Econom\'ia y Competitividad,
the Conselleria d'Economia i Competitivitat and Conselleria d'Educaci\'o, Cultura i Universitats of the Govern de les Illes Balears,
the National Science Centre of Poland,
the European Commission,
the Royal Society, 
the Scottish Funding Council, 
the Scottish Universities Physics Alliance, 
the Hungarian Scientific Research Fund (OTKA),
the Lyon Institute of Origins (LIO),
the National Research Foundation of Korea,
Industry Canada and the Province of Ontario through the Ministry of Economic Development and Innovation, 
the Natural Science and Engineering Research Council Canada,
Canadian Institute for Advanced Research,
the Brazilian Ministry of Science, Technology, and Innovation,
Russian Foundation for Basic Research,
the Leverhulme Trust, 
the Research Corporation, 
Ministry of Science and Technology (MOST), Taiwan
and
the Kavli Foundation.
The authors gratefully acknowledge the support of the NSF, STFC, MPS, INFN, CNRS and the
State of Niedersachsen/Germany for provision of computational resources.
This article has been assigned the document number \href{https://dcc.ligo.org/LIGO-P1500217/public/main}{LIGO-P1500217}.